# Macroscopic Facilitation of Glassy Relaxation Kinetics: Ultra Stable Glass Films with Front-Like Thermal Response.

Sébastien Léonard[1,2] and Peter Harrowell[1]

[1]*School of Chemistry,*
*University of Sydney, Sydney NSW 2006 Australia*

[2]*Laboratoire Matière et Systèmes Complexes (MSC), UMR 7057,*
*CNRS-Université Paris-Diderot, F-75205 Paris Cedex 13, France*

**Abstract**

The recent experimental fabrication of ultra stable glass films via vapour deposition [Science **315**, 353 (2007)] and the observation of front-like response to the annealing of these films [Phys.Rev.Lett. **102**, 065503 (2009)], have raised important questions about the possibility of manipulating the properties of glass films and addressing fundamental questions about kinetics and thermodynamics of amorphous materials. Central to both of these issues is the need to establish the essential physics that governs the formation of the ultra-stable films and their subsequent response. In this paper we demonstrate that a generic model of glassy dynamics can account for the formation of ultra-stable films, the associated enhancement of relaxation rates by a factor of $10^5$, the observation of front-like response to temperature jumps and the temperature dependence of the front velocity

e-mail: sebastien.leonard@univ-paris-diderot.fr, peter@chem.usyd.edu.au

PACS:  64.70.qj, 64.70.kj

## 1. Introduction

Molecules at the glass-vapour interface are more mobile than those in the bulk glass. Extensive experimental evidence exists for this kinetic enhancement from both free-standing and supported polymer films [1-3]. While a crystal must typically be within ~ 2K of its melting point before undergoing surface melting [4], glasses can apparently exhibit liquid-like kinetics at their surface at 20K or more below the bulk glass transition temperature [1-3]. This phenomenon is of obvious relevance to surface treatments of glasses. Perhaps the most remarkable consequences of surface-enhanced dynamics have only recently been demonstrated. In 2007, Swallen et al [5-7] reported that films of organic glasses, comprised of 1,3,5-(tris)-napthylbenzene (TNB) or indomethacin (IMC), had been prepared with significantly lower enthalpies and higher densities [5] than observed in slow cooling of bulk samples. How much more stable the deposited films are can perhaps be best appreciated in terms of the enthalpies obtained, the lowest of which, for TNB, corresponded to the equilibrated value at a temperature ~ 40K *below* the bulk $T_g$. This is a remarkable result. Previous reports [8] on the vapour deposition of amorphous materials (at deposition temperatures < $0.7T_g$) had given no indication of such stability, finding instead the resulting films to be unstable, exhibiting large exotherms on warming. In contrast, the authors of ref. 7 report that films deposited at a higher temperature, ~ $0.9 T_g$, can be extremely stable. They estimated that to achieve a similar enthalpy through the aging of a bulk sample would take roughly $10^5$ times longer than required for the vapour deposition process. A second striking feature of these very stable films is the manner in which they respond to a jump in temperature to a value above the bulk $T_g$. Instead of the relaxation to the new temperature occurring uniformly throughout the sample, Swallen et al [9] found that the relaxation proceeded via a propagating front initiated at the free surface and, sometimes, another front starting from the substrate.

The picture of an interface-mediated growth of a liquid state into the glass is striking and invites a range of interpretations. In this paper we present the results of modelling the experiments of the Ediger group using a minimalist model of cooperative dynamics, the facilitated kinetic Ising model [10], suggested by Fredrickson and Andersen as a model of the collective relaxation in a glass forming liquid. We shall show that all of the features reported by Ediger and coworkers can be qualitatively accounted for in a model designed to represent the cooperative dynamics of a supercooled liquid without the involvement of any equilibrium phase transition. One conclusion from our work is that the glass-vapour interface can provide an accessible macroscopic window onto kinetic and structural features of the dynamic heterogeneities that dominate glassy relaxation on molecular length scales.

## 2. The Experimental Background

In 2007 Swallen, Kearns, Ediger and coworkers [5-7] demonstrated that amorphous films of the glass-formers tris(napthylbenzene) (TNB) and indomethacin (IMC) formed by vapour deposition below $T_g$ had achieved considerably lower enthalpies and higher densities than observed during slow cooling of the bulk melts. Analogous results have subsequently been reported for smaller molecules by Ishii et al [15,16] and Léon-Gutierrez et al [17,18]. This stability is in sharp contrast to the very unstable films formed by deposition at lower temperatures [8]. At the lowest deposition rate (0.15 nm/s ), a TNB film was produced [7] with a fictive temperature $T_f \sim 307K$, that is $\sim 40K$ *below* the bulk $T_g$ (i.e. $\sim 0.88 T_g$). (Here, the fictive temperature corresponds to the temperature one would require to achieve the same enthalpic state at equilibrium.) The increased stabilization has been attributed to an enhanced relaxation rate in the deposited films. To estimate the magnitude of this kinetic enhancement, Kearns et al [7] have measured the decrease of $T_f$ in bulk IMC as a function of aging time and compared this with the dependence of the film $T_f$ on a surface time defined as 1 nm/deposition rate (nm/s). The two curves of $T_f$ vs log (t) (which, interestingly, exhibit similar slopes) correspond to time scales that differ by a remarkable factor of $10^7$, the factor by which the relaxation kinetics has been enhanced during deposition.

Fixing the deposition rate, Swallen et al [19] established that the relaxation time of a TNB film (and, by extension, its stability) has a maximum at a deposition temperature $\sim 0.85\ T_g$. Lowering the deposition temperature to $\sim 0.7\ T_g$ resulted in a rapid drop in the relaxation time as vapour deposition, recovering the unstable films found in earlier experiments. The striking narrowness of the peak in relaxation times helps explain why these stable films had escaped observation up till now. While these ultra-stable films are disordered, they exhibit a number of features that distinguish them from the amorphous states found in the bulk. The heat capacity peak observed on heating the low $T_f$ film can exhibit a split peak [7]. Dawson et al [20] have recently reported the appearance of a new scattering peak at a wave vector roughly half that of the first peak in the bulk structure factor in films of IMC deposited below $T_g$-20K.

The relaxation of the ultra-stable films following a jump in temperature to above $T_g$ was found [9] to proceed via a front of the liquid-like phase that propagated in from the free surface. The relaxation front was found to move with a constant velocity (no transient lag was observed) which varied with the final temperature in a manner proportional to the equilibrium diffusion constant at that temperature. In some cases, the film adjacent to the substrate also acted as an initiation site for the relaxation to the liquid state during annealing. One puzzling observation [9,21] is that the liquid produced in the wake of the propagating front from the free surface exhibits a diffusion constant roughly twice that of the equilibrium liquid. This enhanced mobility persists for times considerably longer than the equilibrium structural relaxation time of the liquid at this temperature.

### 3. The Model

The facilitated kinetic Ising model was introduced in 1984 [10] as a 'simplest' model of cooperative dynamics. The spins are intended to model some aspect of local structure that determines whether dynamics can occur or not. There is no interaction between spins and, hence, no equilibrium phase transition. Spins do, however, exert a kinetic influence on their neighbours. A spin cannot flip unless *n* or more of its neighbours are in the 'up' state. An applied field with strength *H* favours down spins and so, as the temperature T (in reduced units of $H/k_B$) decreases, the fraction of up spins diminishes and relaxation becomes increasingly dominated by the rare regions in which relaxation is still permitted. While the name 'facilitated' suggests kinetic assistance, it would, in retrospect, be more accurate to describe this model as capturing 'structurally determined' kinetics. An attraction of this model is that it provides an explicit link between, on the one hand, the macroscopic phenomenology of relaxation functions and their temperature dependence and, on the other hand, the microscopic fluctuations of dynamic heterogeneities [11,12,13]. The facilitated spin model has become something of an archetype of the character and evolution of dynamic heterogeneities as the glass transition is approached.

In this study we shall use the 3-spin facilitated Ising model on a cubic lattice [22] (referred to in this paper as the 3D-3spin model). The model has been studied numerically by Graham et al [23]. The constraint that a spin cannot flip unless it has at least 3 neighbours in the 'up' state is an onerous one and produces a very rapid increase in the relaxation time on cooling. Graham et al fitted the temperature dependence of the relaxation time $\tau$ to the Volger-Fulcher expression $\tau = \exp(A/[T-T_o])$ with a value of $T_o = 0.68$. This corresponds to an apparent divergence of $\tau$ at an up spin concentration $c = 0.187$, considerably higher (as we shall see) than the more extensively studied 2D and 1D models.

The equilibrium concentration of up spins is given by

$$c = \frac{1}{1+e^{1/T}} \tag{1}$$

The fictive temperature $T_f$ is obtained by inverting this relation, i.e.

$$T_f = \frac{1}{\ln(c^{-1}-1)} \tag{2}$$

and is just equivalent to specifying the concentration $c$ of up spins. At equilibrium, $T_f = T$. To generate a configuration consistent with the equilibrium state we simply select a random distribution of up spins, the concentration of which is given by Eq. 1. The subtle point is to establish that the configuration so generated is kinetically accessible from a high temperature state. In order for a configuration to be kinetically accessible from a high temperature quench it must belong to the partition of configuration space to which the all up state belongs [22]. To ensure that a random spin configuration belongs to the high temperature partition we test it by flipping every spin that can flip up until no further spins can be flipped. If this final state is the all up state then our original configuration was a member of the high temperature partition, otherwise, the initial configuration is discarded and another random configuration is generated and the test repeated. By this method we can produce the correct equilibrium distribution of configurations at temperatures too low to equilibrate dynamically in any reasonable time. (Using this algorithm, the problem of the high temperature partition becomes essentially the same as that of bootstrap percolation [22, 24].)

In 1991 we used a facilitated kinetic Ising model [7] to study surface enhanced relaxation dynamics. The vapour side of the vapour–glass interface was modelled as a layer of permanent up spins. (We shall extend this earlier idea in this paper through the inclusion, in the model, of layer-by-layer deposition.) It is important to emphasise the point that, in this treatment of glass surfaces, the enhanced relaxation rates is achieved through the application of the same kinetic constraints that give rise to the dynamic heterogeneities in the bulk phase. The only difference between the two phenomena is that, in the free surface, we have, effectively, a macroscopic dynamic heterogeneity. The attraction of explaining the surface behaviour using the same model used to account for bulk dynamic heterogeneities is not just one of economy. If the surface enhanced dynamics is a macroscopic manifestation of the kinetic fluctuations responsible for the microscopic heterogeneities then the study of the surface effect can provide us with a valuable window onto the behaviour of dynamic heterogeneities.

The vapour side of the free surface and the solid substrate are modelled as a layer of spins that are permanently all up or all down, respectively [14]. (A kinetically more complex substrate can be modelled with some distribution of up and down spins.) A spin adjacent to either interface will see a different average environment to that sampled in the bulk. In this paper we shall work with the (100) plane, the surface that, as explained in the Appendix, produces the *minimum* surface enhancement to kinetics for this model.

For the dynamics of the spin model we use the Metropolis Monte-Carlo (MC) algorithm, supplemented with the kinetic constraints. We have used a continuous time Monte-Carlo algorithm described in ref. 25. Time is expressed in terms of Monte Carlo cycles where one cycle is equal to N attempted spin flips where N is the number of sites in the glass-forming film. Bulk simulations are performed with periodic boundary conditions in all three directions while, in the film, periodic boundaries are only employed in the two directions orthogonal to the surface normal.

To model the layer-by-layer deposition we shall incrementally increase the number of layers in the film from the substrate as sketched in Fig. 1. The upper layer is always in contact with the free surface boundary. The initial configuration of each new layer is the all up state. When the average attempt per spin in the current surface layer equals $1/\gamma$ ($\gamma$ being the deposition rate), we allow the next layer to start relaxing. This protocol allows each surface layer to achieve the same degree of relaxation during the deposition process. We continue this deposition until the required number of layers is reached. Annealing of the film after deposition is a simpler process, requiring only that the temperature be set to the annealing temperature and the subsequent dynamics monitored.

**4. Surface Enhancement of Relaxation Kinetics and Reduction of Fragility**

As already noted, there is considerable evidence of enhanced kinetics at the glass-vapour interface. In an elegant experiment involving the relaxation of surface indentations, Fakhraai and Forrest [26] measured the temperature dependence of the relaxation time at the surface of polystyrene. They found that the surface relaxation was not only enhanced relative to that of the bulk but that it was considerably less fragile. At sufficiently low temperatures, the relaxation became almost independent of temperature. While surface enhancement of kinetics does not *require* a decrease in fragility, such a decrease will ensure that the gap between surface and bulk relaxation rates increases rapidly on cooling and, hence, promote the formation of ultra stable films. Stevenson and Wolynes [27] have presented a theory of surface enhancement of kinetics in which the volume of the activation region at the surface is halved resulting in a simple relation between the surface relaxation time $\tau_S$ and the relaxation time $\tau$ of the bulk, i.e.

$$\tau_S \propto \sqrt{\tau} \tag{3}$$

This result is an example of enhanced surface kinetics without a change in fragility.

The free surface of the 3D-3spin model relaxes more quickly than the bulk because the presence of a layer of permanent all up spins on the vapour side of the interface ensures that the local condition for relaxation has been partially met for every surface site. Stated more generally, we explain the surface enhanced kinetics as resulting from the fact that the kinetics at the surface suffers less constraint than in the bulk. This is the only physical role played by the surface in this model. We do not invoke specific changes in molecular structure near the surface nor particular surface fluctuations. The degree to which this simple treatment is successful in reproducing the observed behaviour will provide some insight into whether greater microscopic detail is required in the surface modelling.

Consider now the (100) layer constituting the surface layer of our film (see Fig. 2). Above this layer, we have the vapour (or vacuum) phase, characterised by a layer of all up spins (i.e. no source of constraint). That means that every spin in our surface layer has, by construction, at least one up neighbour and, therefore, only needs to find two more to enable a spin flip. If we were to treat the layer beneath the surface layer as being all down (and, in doing so, certainly underestimate the up spin concentration there), then relaxation in the surface layer would correspond to that of a 2D-2spin model on a square lattice. The 2D-2spin model represents a rigorous upper bound on the surface relaxation time, since the (neglected) contribution of the up spins in the layer beneath will only serve to speed up relaxation. The surface enhancement in kinetics can, therefore, be reasonably captured by the comparison of the relaxation times of the 3D-3spin model (i.e. the bulk) with the relaxation time of the 2D-2spin model at the same temperature. Previously [11], we have obtained the following expression for the relaxation time of the 2D-2spin relaxation time,

$$\tau_{2D} = \tau_{0D} 0.34 Q_{2D}^{-3.8} \tag{4}$$

where

$$Q_{2D} = \left[6c^2(1-c)^2 + 4c^3(1-c) + c^4\right]\exp\left(\frac{(1-c)^8}{8\ln(1-c)}\right) \tag{5}$$

and $\tau_{0D}$ is the elementary time scale for a flip attempt. Following the same argument used in ref. 11, we have obtained an analogous expression for the relaxation time of the 3D-3spin model (see Appendix for details including the comparison of Eq.5 with simulation data), i.e.

$$\tau_{3D} = \tau_{0D} 31.77 Q_{3D}^{-3.95} \tag{6}$$

where

$$Q_{3D} = q \exp\left[-6(1-c)^7 \sqrt{\frac{-1}{24\ln(1-c)}} \left[1 - \Phi(\sqrt{-6\ln(1-c)})\right]\right], \tag{7}$$

$$q = 120c^3(1-c)^3 + 30c^4(1-c)^2 + 6c^5(1-c) + c^6 \tag{8}$$

and $\Phi(x)$ is the error function,

$$\Phi(x) = \frac{2}{\sqrt{\pi}} \int_0^x dt\, e^{-t^2}$$

In Fig. 3 we have plotted the relaxation times for the surface and bulk using Eqs.4 and 6, respectively. The enhancement of the surface kinetics is dramatic. The bulk relaxation time does not exhibit a divergence at T = 0.68, as suggested by the Volger-Fulcher fit reported in ref. 23. If we (arbitrarily) define the glass transition temperature as corresponding to a relaxation time of $10^{10}$, then the bulk $T_g$ is 0.46. At this temperature, the surface relaxation rate is enhanced over that of the bulk by a factor of ~ $10^5$. From Fig.3 we see that the surface would only achieve a relaxation time of $10^{10}$ at T ~ 0.32.

The bulk and surface relaxation times also exhibit different temperature dependencies as characterised by the fragility *m*, defined by [28]

$$m = \left.\frac{d\log(\tau)}{d(T_g/T)}\right|_{T=T_g} \tag{6}$$

Again using Eqs. 4 and 6 (and, for the $T_g$'s the temperature at which the relaxation time is $10^{10}$) we find the surface and bulk fragilities to be 26 and 18, respectively. We see that the surface is less fragile than the bulk, consistent with experiments [26], but the difference is only modest (TNB, for comparison, has m = 66 [28]). In fact, the relationship, Eq. 3, proposed by Stevenson and Wolynes [27] actually works quite well here to describe the difference between the bulk and (100) surface relaxation times (see Fig.3). This is surprising since it is not clear that an activation complex argument (employed in ref. 27) applies to the facilitated model. In particular, where the authors of ref. 27 regard the surface as reducing the activation volume at the surface by half (the origin of the ½ exponent in Eq.3), the (100) surface of the 3D-3spin facilitated kinetic Ising model only effects 1/6 of the surface spins nearest neighbours.

## 5. The Stability of Deposited Films

Next, we shall consider the stability of the deposited films as a function of deposition temperature and deposition rate. To do so we shall use the fictive temperature defined in Eq. 2. For readers uncomfortable with things fictive, note, with reference to Eq.2, that we could equivalently measure stability by the up spin concentration. In Fig.4 we plot the fictive temperature $T_f$ as a function of the deposition temperature, T, for a 20 layers film created using a number of different deposition rates. The equilibrium line, $T_f = T$, is included for reference. For the bulk quench, 20 layers were cooled using periodic boundary conditions in all directions (i.e. with no surfaces present). The sample was then held at the deposition temperature for a time period equal to that of the total time required to deposit 20 layers at a rate $\gamma = 10^{-3}$.

We find that all of the simulated curves exhibit a minimum with respect to the deposition temperature. For a given deposition rate, there is a temperature below which relaxation can no longer keep up rate of deposition and an increasing number of high energy configurations are trapped. The temperature and energy associated with this minimum decreases with decreasing rate of deposition. The deposited films are clearly achieving a considerably lower energy than the analogous bulk quench.

Each of the curves in Fig.4 can be characterized by three quantities (see Fig.5 for graphical definitions) :

- $T_{sg}(\gamma)$, the value of the deposition temperature at which $T_f$ departs from the equilibrium condition,

- min $T_f(\gamma)$ and the corresponding $T_{min}(\gamma)$, the minimum fictive temperature obtained for a given deposition rate and the corresponding value of the deposition temperature at which this optimally stable film was obtained, and

- $T_f(0)$, the value of $T_f$ when the film is deposited at T = 0.

From Fig.4 we find that $T_f(0)$ ~1.6, independent of the how the film is generated (or even if it is a film at all). This value of $T_f$ corresponds to an up spin concentration of ~ 0.35. This is the concentration of spins remaining after all the flippable spins have been eliminated by a process of randomly flipping flippable spins down. In packing terminology, this state is roughly analogous to random loose packing.

Of greater physical significance is the temperature $T_{sg}(\gamma)$ at which $T_f$ departs from the deposition temperature T. This is the analogue, for the vapour deposited film, of the glass transition temperature $T_g$ and provides a similar useful reference temperature with regards the material properties of the film. Where, above, we arbitrarily chose a relaxation time of $10^{10}$ to illustrate how the surface glass transition was depressed relative to the bulk, here we acknowledge that the deposition rate *imposes* a time scale on the surface and that the corresponding $T_{sg}$ provides the measure of the surface glass transition most relevant to vapour deposition. Note that $T_{sg}(\gamma)$ ~ min $T_f(\gamma)$, the lowest fictive temperature available at a given deposition rate (see Figs. 2 and 3). Since equilibration of the deposited film takes place at the interface, min $T_{sg}(\gamma)$ corresponds to the deposition temperature at which the deposition rate $\gamma$ exceeds the surface relaxation rate $1/\tau_s$, i.e.

$$\gamma \tau_s(T_{sg}) \sim 1 \tag{3}$$

To the degree that $T_{sg}$ determines the lowest fictive temperature achievable, the depth that we can descend into an energy landscape will depend on how low we can depress the surface glass transition.

The minimum fictive temperature, min $T_f$, we have achieved is 0.39 (using a deposition rate of $10^{-4}$ at T = 0.2). To reach a fictive temperature of 0.39 via the cooling of a bulk sample would take, at least, $10^6$ times longer than the deposition simulations (estimated using Eq. 6). The deposition algorithm, therefore, represents a speed up similar to that estimated for the TNB films [7]. This demonstration of an algorithm that makes accessible low energy regions of the configuration space is noteworthy and raises the possibility of new algorithms which exploit speed up's similar to that afforded by the free surface and can be developed as general tools in the study of glass forming liquids.

In light of the highly heterogeneous route by which the deposited films were produced, it is of interest to ask how homogeneous are the resulting films? We find, not surprisingly, that the concentration c of up spins is uniform except for the layers adjacent to the substrate. The up spin concentration (and, hence, the fictive temperature) is systematically higher in the layer directly adjacent to the substrate. This is a result of the choice of the substrate as a 'mobility reducing' surface (i.e. as an all down spin region). As the first layer is deposited, the kinetic enhancement of the free surface is being, to some extent, countered by the kinetic suppression exercised by the substrate. Hence, we are seeing poorer equilibration at the substrate than in the rest of the film. Anisotropies in the film, if present, would show up in terms of spin-spin correlations. We shall leave to a future paper consideration of anisotropies in such correlations in the deposited film.

The layer-by-layer distribution of relaxation kinetics of the deposited film provides a complimentary picture to that of the structure. In Fig. 6 we plot the relaxation time $\tau_i$ for the ith layer defined [14] as

$$\tau_i = \int_0^\infty dt F_i(t) \qquad (4)$$

where $F_i(t)$ is the average fraction of spins that have not yet flipped within time t in layer i. To avoid the problem posed by non-stationarity of a nonequilibrium film, the spatial distribution of $\{\tau_i$'s$\}$ plotted in Fig.6 were obtained from equilibrated films. Despite the higher spin density observed adjacent to the substrate we find that relaxation is still slower against the substrate due to the spin down condition employed. The layer adjacent to the free surface, in contrast, has a relaxation time is roughly two and half orders of magnitude smaller than in the bulk. The decay of the layer relaxation time as one moves from the free surface into the bulk, as shown in Fig. 6, defines a kinetic correlation length [14].

**5. Bulk Relaxation vs Front Propagation Following a Temperature Jump**

The response of a glass forming material to a temperature jump represents a standard feature of the phenomenology by which glassy behaviour is characterised. Nonlinearity in the transient response is common. The rate of equilibration of a sample annealed at a temperature T is typically faster if it follows a temperature *drop* of some ΔT than a temperature *jump* by the same ΔT [29]. Glassy response to a temperature jump can even show a non-monotonic (overshoot) in state variables, a feature often referred to as the Kovacs effect [30]. Both the asymmetry in the sign of ΔT [29] and the Kovacs effect [31] have previously been reproduced using facilitated Ising models.

Swallen et al [9] have established that the ultra stable TNB film exhibits a number of interesting new features when relaxing after a temperature jump. These features are:
i) annealing can either take place homogenously throughout the film or, if the deposition temperature is low enough, annealing is in the form of a propagating front starting from the free surface
ii) the annealing front (from the free surface) exhibits a constant velocity with no observable induction time, and
iii) the velocity of the annealing front scales with annealing temperature in the same manner as the diffusion constant of the liquid equilibrated at the annealing temperature. Again, we are referring to the front initiated at the free surface.

Turning to the facilitated model, we shall compare the response under annealing at T = 0.8 of three films – all deposited at the same rate, $\gamma = 10^{-3}$ but with different deposition temperatures, T = 0.3, 0.5 and 0.7. In Fig. 7 we plot the evolution of the system in the form the spatial profiles of the persistence of spin states for a sequence of times. The persistence of a layer is the fraction of spins that have not flipped within the time since the temperature jump. The persistence drops to zero quickly at the layer adjacent to the vapour (layer 20) due to the enhanced mobility at the surface. In the film deposited at T = 0.3, the persistence decreases with time as a propagating front moving at a constant velocity. For a film deposited at a higher temperature, T = 0.7, we find a completely different behaviour. While the diminished persistence at the free surface is still evident, the persistence decreases uniformly throughout the sample with little influence from the surface. This behaviour indicates that the bulk relaxation is now the major pathway for film relaxation. The crossover between front and bulk relaxation can be seen at the intermediate deposition temperature of T = 0.5. We do not see any front propagating from the substrate interface.

We have carried out annealing calculations for a wide range of deposition temperatures and deposition rates. The results are summarised in the non-equilibrium phase diagram plotted in Fig. 8. We have plotted the fictive temperature $T_f$ vs the deposition temperature T. (Since, for a given deposition temperature, the fictive temperature is a monotonically increasing function of the deposition rate, one can regard Fig. 8 as the phase diagram in the space of deposition control parameters.) The outcome of each annealing run is placed into one of three categorise: bulk-like relaxation, front propagation and a crossover behaviour between bulk relaxation and front propagation, and the ($T_f$, T) half plane is divided accordingly. We find that the boundary of the front propagation region is reasonably well described by a constant fictive temperature of $T_f \sim 0.6$. This is a non-trivial result as this space includes a range of deposition temperatures and rates. The crossover region is narrow for the 20 layer film but we note that this region must extend to lower fictive temperatures with increasing film thickness. This size dependence is a consequence of the fact that the time required for a moving front to relax a film is proportional to the film thickness. The longer the front-mediated relaxation time, the slower the rate of bulk relaxation necessary to kinetically compete with front growth. The linear bulk relaxation is observed above the bulk glass transition temperature and below this temperature for systems very close to equilibrium.

**6. Dynamics of the Propagating Front: Position and Width**

As shown in Fig. 7, we can characterise the relaxation front by a profile of the average spin through the interface. We shall define the position $z$ of the interface to be that point along the surface normal corresponding to a total spin (proportional to the film energy) half way between the final and initial values. The width $w$ of the interface is defined as the distance, again parallel to the surface normal, between 0.9 and 0.1 of the difference between the persistence before and behind the front. The position $z$ the interface is plotted as a function of time in Fig. 9 for a film deposited at T = 0.3 and then subjected to annealing at a range of temperatures. We find that $z$ increases linearly in time with a velocity that increases with the annealing temperature. We find no sign of an initial induction time in which the front is stationary. For annealing temperatures close to the bulk glass temperature (T = 0.7) there appears to be an initially faster front velocity. This transient behaviour corresponds to a surface reorganization that has become observable due to the low front velocity.

For an interface associated with a first order phase transition, the moving front attains a steady state width, its shape constrained by the free energy that governs its growth. In contrast, front growth associated with unrelaxed surface structure such as occurs in models of non-equilibrium growth such as aggregation and liquid percolation through porous materials [32] typically exhibits widths that grow with time. It is common to assume [33] that the time evolution of the width $w$ can be written as $w \sim t^\beta$ where $\beta$ is called the growth exponent. The standard growth models for a first order transition produce a fixed interface width (i.e. $\beta = 0$). In contrast, an example of strong kinetic roughening such as random deposition without relaxation [33] exhibits a growth exponent $\beta = 0.5$. It is striking, therefore, that we find the annealing fronts for the facilitated kinetic Ising model exhibit a larger growth exponent with $\beta \sim 1$ as shown in Fig. 10. A linear increase of the width with time is consistent with different parts of the interface exhibiting long lived differences in growth rate. It could also result from an accumulation of inherent fluctuations in 'flipped' spins in the film. The absence of a surface energy between the high and low $T_f$ states means that there is no tendency to 'smooth' out the high q structure of the interface. The growth exponent clearly provides a striking distinction between different interpretations of the propagating front – i.e. phase transition vs nonlinear kinetics. The experimental growth exponents for the ultra stable amorphous films have not yet been determined.

What is the connection between the roughness of the propagating front and the dynamic heterogeneities of the amorphous film into which it advances? In Fig.11 we show images of the projection of the local spin dynamics onto a plane which includes the surface normal for a sequence of 'snapshots' of a propagating front. Each image represents the spatial distribution of the persistence of spins at a sequence of times during the propagation of the front. If any site within three layers of the xz plane has flipped at least once it is coloured white, otherwise it is left black. We see that inherent dynamic heterogeneities occur throughout the film but are typically trapped and unable to propagate. Initially we do not see any sign that the interface has any more flippable sites than the bulk. What does differentiate the surface layer is that the probability of trapping of any flippable sites is essentially zero. This means that sites that can flip there are more likely to able to go on and propagate that relaxation, especially laterally. By this means the relaxation front is advanced into the bulk of the film. Of particular interest is the behaviour of the film near the crossover between front propagation and bulk relaxation. As already pointed out, in this crossover region relaxation can be described occurring via a front whose width grows rapidly. The snapshots in Fig.11 provide explicit evidence that the broadening is a direct consequence of the front 'accumulating' the localised pockets of relaxation inherent in the dynamic heterogeneities of the bulk film.

In ref. [9], Swallen et al reported on the dependence of the front velocity on the annealing temperature. They found that the velocity was proportional to the equilibrium diffusion constant in the bulk liquid at the annealing temperature. We have looked at a similar correlation in the facilitated kinetic Ising model, this time between the front velocity v and the equilibrium relaxation time τ (defined as the decay time of the persistence function [11] ) in the bulk system at the annealing temperature. As shown in Fig. 12, we find the relation is well described by a power law over the range of annealing temperatures considered, with

$$v \sim 0.0033 \, \tau^{-0.83} \qquad (4)$$

Through this relation, the temperature dependence of the front propagation rate is determined by that of the liquid relaxation as observed experimentally.

The transition between front propagation and bulk relaxation provides information about the stability of the amorphous film that is directly accessible to observation. The elegant experiments of Ediger and coworkers are quite involved – requiring the synthesis of deuterated material, vacuum deposition and secondary ion mass spectrometry. It would, therefore, be useful if the mechanism of response to a temperature jump could be established without the need to explicitly resolve the evolving structure. The change in the mechanism of relaxation as the deposition temperature is decreased should show up in the kinetics of the total spin (or energy) of the film. In Fig. 13 we plot the time dependence of the concentration of up spins in films formed at a range of temperatures when annealed as T = 0.8. The transition between front propagation and bulk relaxation is clear. The observation of relaxation proceeding linearly with time provides evidence of front propagation without having to undertake the difficult experiments carried out by Ediger and coworkers. Furthermore, the front velocity is simply the slope of this linear increase in relaxation. Kearns et al [34] have recently employed this approach using calorimetry to establish which relaxation mechanism is dominant in a film. In an interesting development, the control variable in ref. 34 was not deposition temperature or rate but the thickness of the film.

Finally, we shall make some observations about the effect of the surface on aging of a vapour deposited film at sufficiently low $T_f$ such that the surface relaxation time is significantly faster than that of the bulk (i.e. a film that would respond to a temperature jump via front propagation). Aging refers to simply holding the film at a low temperature for a long time. In the facilitated model, the surface influence on aging is self limiting. As the surface relaxes more quickly, the up spin concentration drops more quickly than in the bulk until the point is reached where the difference between the surface and bulk relaxation times has vanished. From this point on, the film relaxes just as a bulk sample with the lower concentration of up spins at the surface, arising from the faster relaxation, only penetrating one layer in to the film.

## 7. Conclusions

We have shown that a simple model of glassy relaxation in which relaxation is dominated by the local fluctuations in constraints is capable of reproducing the following experimental observations from vapour-deposited amorphous films:

i) fictive temperatures well below the glass transition temperature, characterising the ultra-stable glass,
ii) a surface enhanced relaxation kinetics that sees the ratio $\tau_{bulk}/\tau_{surface}$ exceed $10^5$ in the most stable films,
iii) the presence of a minimum in $T_f$ as a function of the deposition temperature, with a rapid rise in the film energy as the deposition temperature is decreased below the optimal value
iv) front-like propagation of relaxation during annealing at a temperature above $T_g$ for films prepared below a critical fictive temperature,
v) an annealing front that propagates without any significant induction time and at a constant velocity, and
vi) a front velocity that scales roughly linearly with the equilibrium relaxation time at the annealing temperature.

We conclude that the facilitated kinetic Ising model does indeed provide a reasonable and consistent account of the experimental situation (both in terms of the underlying physics and the resulting phenomena). The proposition that the enhanced surface dynamics and its resulting consequences (ultra-stable glasses and propagating annealing fronts) are direct results of the same physics responsible for dynamic heterogeneities is significant. From this perspective, the vapour-glass interface behaves as a macroscopic dynamic heterogeneity. The surface enhanced kinetics is related to the analogous enhancement found at the microscopic level in the mobile regions generated by thermal fluctuations. The evolution of the surface kinetics as a result of a temperature jump obeys the same rules as does the response of the microscopic fluctuations to a similar perturbation. The experimental observable that is, perhaps, the most informative concerning the character of the microscopic heterogeneities is the magnitude and temperature dependence of the surface relaxation time (including the surface glass transition temperature $T_{sg}$). Our results suggest that observations of the time dependence of the front width could provide valuable clues as to the length scales of the intrinsic heterogeneities in the film.

By construction, the facilitated kinetic Ising model does not have any thermodynamic phase transition. While not excluding the possibility that liquid-liquid phase transitions might be involved in the behaviour of the deposited films, the success of the spin model in describing the deposition experiments demonstrates that such phase transitions are not necessary to explain observations such as front propagation. The results of this paper strongly support the proposal that all that is needed to account for the production of ultra-stable amorphous films through deposition and their subsequent relaxation via front propagation is to have a material that, on supercooling, exhibits dynamic heterogeneities.

The observation of propagation fronts in a model of collective relaxation like the facilitated Ising model underscores the essential characteristic of this model of such relaxation and, presumably, the characteristic of the class of kinetics to which this model belongs. Perhaps the most useful description of this kinetics class is *random threshold growth* [35]. In 1991, Butler and Harrowell [11] pointed out that the relaxation in the 2D-2spin facilitated kinetic Ising model corresponded to growth from clusters of flippable spins, only a few of which were so configured so as to escape the ultimate trapping of the growth. The results of this present paper can be summarised by the statement that the free surface represents an extended cluster of flippable spins so large as to *always* (except at very low T) escape this trapping of growth. Independently, Wolynes [36] has presented a description of the propagation of rejuvenation (i.e. an up temperature anneal) in glass forming material as an autocatalytic front, analogous to that of a flame.

An important unresolved question is whether the ultra stable films produced by vapour deposition are the same as would have been generated by extremely long aging of the bulk sample or are they different, possibly with an anisotropy that reflects the anisotropic manner of their formation. The appearance of a new scattering peak at small wavevector in IMC [20] and, in the case of indomethacin, the appearance of two peaks in the differential scanning calorimetry curves certainly point to something new in the real films. We will report elsewhere [37] on a detailed study of differences between the vapour deposited films and isotropically generated films in the facilitated kinetic Ising model.

**Acknowledgements**

We gratefully acknowledge the many enjoyable conversations with Mark Ediger, Stephen Swallen, Ken Kearns and Kevin Dawson which inspired and guided this research. The work has been funded by a Discovery grant from the Australian Research Council.

**References**

1. R. C. Bell, H. F. Wang, M. J. Iedema, and J. P. Cowin, J. Am. Chem. Soc. **125**, 5176 (2003).
2. C. J. Ellison and J. M. Torkelson, Nat. Mater. **2**, 695 (2003).
3. J. S. Sharp and J. A. Forrest, Phys. Rev. Lett. **91**, 235701 (2003).
4. J. G. Dash, Contemp. Phys. **30**, 89 (1989); *ibid*, **43**, 427 (2002).
5. S.F.Swallen, K. L. Kearns, M.K.Mapes, Y.S.Kim, R.J.McMahon, M.D.Ediger, T. Wu, L. Yu and S. Satija, Science **315**, 353 (2007)
6. K. L. Kearns, S.F.Swallen, M.D.Ediger, T.Wu and L.Yu, J. Chem. Phys. **127**, 154702 (2007)
7. K.L.Kearns, S.F.Swallen, M.D.Ediger, T.Wu, Y.Sun and L.Yu, J. Phys. Chem. B **112**, 4934 (2008).
8. K. Ishii and H. Nakayama, J. Non-Cryst. Solids **353**, 1279 (2007).
9. S.F.Swallen, K. Traynor, R.J.McMahon, M.D.Ediger and T. E. Mates, Phys.Rev.Lett. **102**, 065503 (2009)
10. G. H. Fredrickson and H. C. Andersen, Phys. Rev. Lett. **53**, 1244 (1984).
11. S. Butler and P. Harrowell, J. Chem. Phys. **95**, 4454 (1991).
12. M. Foley and P. Harrowell, J. Chem. Phys. **98**, 5069 (1993).

13. J. P. Garrahan and and D. Chandler, Phys. Rev. Lett. **89**, 035704 (2002).
14. S. Butler and P. Harrowell, J. Chem. Phys. **95**, 4466 (1991).
15. K. Ishii, H. Nakayama, S. Hirabayashi and R. Moriyama, Chem. Phys. Lett. **459**, 109 (2008)
16. K. Ishii, H. Nakayama, R. Moriyama and Y. Yokoyama, Bull. Chem. Soc. Jpn. **82**, 1240 (2009).
17. E. Léon-Gutierrez, G. Garcia, A.F.Lopeandia, J.Fraxedas, M.T.Clavaguera-Mora and J. Rodriguez-Viejo, JCP **129**, 181101 (2008).
18. E. Léon-Gutierrez, G. Garcia, M.T.Clavaguera-Mora and J. Rodriguez-Viejo, Thermochimica Acta **492**, 51 (2009).
19. S. F. Swallen, K. L. Kearns, S. Satija, K. Traynor, R. J. McMahon and M. D. Ediger, J. Chem. Phys. **128**, 214514 (2008)
20. K. J. Dawson, K. L. Kearns, L. Yu. W. Steffen and M. D. Ediger, Proc. Nat. Acad. Sci. **106**, 15165 (2009)
21. S. F. Swallen, K. Windsor, R. J. McMahon, M.D. Ediger and T. E. Mates, J. Phys. Chem. B **114**, 2635 (2010).
22. G. H. Fredrickson and H. C. Andersen, J. Chem. Phys. **83**, 5822 (1985).
23. I. S. Graham, L. Piche and M. Grant, Phys. Rev. E **55**, 2132 (1997).
24. J. Reiter, J. Chem. Phys. **95**, 544 (1991).
25. A. B. Boratz, M. H. Kalos and J. L. Lebowitz, J. Comp. Phys. **17**, 10 (1975).
26. Z. Fakhraai and J. A. Forrest, Science **319**, 600 (2008).
27. J. D. Stevenson and P. G. Wolynes, J. Chem. Phys. **129**, 234514 (2008).
28. R.Boehmer et al, J.Chem.Phys. **99**, 4201 (1993).
29. G H. Fredrickson and S. A. Brawer, J. Chem. Phys. **84**, 335l (1986); G. H. Fredrickson, Annu. Rev. Phys. Chem. **39**,149 (1988).
30. A. J. Kovacs, Adv. Polym. Sci. **3**, 394 (1963).
31. J.J. Arenzon and M. Sellitto, Eur. Phys. J. B **42**, 542 (2004).
32. M. Sahimi, *Applications of percolation theory* (Taylor & Francis, London, 1994).
33. A-L. Barabási and H. E. Stanley, *Fractal concepts in surface growth* (Cambridge University Press, Cambridge, 1995).
34. K. L. Kearns, M. D. Ediger, H. Huth and C. Schick, J. Phys. Chem. Lett. **1**, 388 (2010); K. L. Kearns, K. R. Whitaker, M. D. Ediger, H. Huth and C. Schick, J. Chem. Phys. **133**, 014702 (2010).
35. T. Bohman and J. Gravner, Rand. Struct. Algorithms. **15**, 93 (1999); L. A. Caffarelli and P. E. Souganidis, Arch. Rational Mech. 195, 1 (2010).
36. P. G. Wolynes, Proc. Nat. Acad. Sci. **106**, 1353 (2008).
37. S. Léonard and P. Harrowell, in preparation.
38. R. J. Glauber, J. Math. Phys. **4**, 294 (1963).
39. S. Whitelam, L. Berthier and J. P. Garrahan, Phys. Rev. E **71**, 026128 (2005).

**FIGURES**

Figure 1. Schematic of the layer-by-layer deposition process as modelled in the facilitated kinetic Ising model. The black cells represent the substrate and the white cells represent the vapour or vacuum. The shaded cells represent the film. A new layer is deposited by allowing the layer of up cells adjacent to the film to relax to the applied field.

Figure 2. Each site (black circle) on the (100) plane at the surface of the film has one neighbour that is up (unfilled circle) by construction. Note that the (100) plane is a square lattice.

Figure 3. A plot of log $\tau$ vs 1/T for the bulk 3D-3spin model and the (100) surface modelled here as the 2D-2spin model. As explained in the text, the 2D-2spin relaxation time provides a rigorous upper bound for the relaxation time for the free (100) surface in the 3D model. The curves are generated using Eqs. 6 and 4, respectively. Eq.4 was obtained from a fit to simulation data in ref. [11]. Eq. 6 is the result of an analogous fit to the 3D data (crosses). The dashed curve is the prediction from ref. 27 for the surface relaxation time $\tau_s$ using $\tau_s = 5.23\sqrt{\tau}$, the value 5.23 was obtained by fitting to the high temperature surface relaxation time.

Figure 4. The fictive temperature $T_f$ of the deposited films (20 layers) as a function of the deposition temperature for different deposition rates. Layer 20 is the free surface (i.e. the layer adjacent to the all up layer). The fictive temperature from a bulk quench is also shown as a function of the annealing temperature. The time scale for the bulk annealing was chosen to equal the annealing time experienced by the first layer deposited in a 20 layer film deposited at a rate of $10^{-3}$. The equilibrium condition $T_f = T$ is indicated by a straight line.

Figure 5. A sketch of the $T_f$ vs T curve (a schematic version of that shown in Fig.4) showing how the quantities $T_{sg}$, min $T_f$ and $T_f(0)$ are defined.

Figure 6. Plot of the average value of the persistence time per layer for layers deposited at the temperatures indicated using a deposition rate of $10^{-3}$.

Figure 7. Relaxation fronts (in the form of persistence profiles) as a function of layer number for a sequence of equally spaced time intervals following the onset of annealing at T = 0.8 for three films deposited at temperatures of a) 0.3, b) 0.5 and c) 0.7, all at a deposition rate of $10^{-3}$. The time intervals between the individual persistence profiles are a) $10^5$, b) 5 x $10^4$ and c) $10^3$.

Figure 8. The nonequilibrium phase diagram indicating whether, when annealed at T = 0.8 after deposition, the film responded with either front propagation, bulk relaxation or some mixed response, corresponding to the cross-over. Lines are drawn to indicate the boundaries between the different regions.

Figure 9. The position of the front as a function of time for a range of annealing temperatures. The film was deposited at T = 0.3 and a deposition rate of $10^{-3}$.

Figure 10. The width of the front vs time for the same examples shown in Fig.9.

Figure 11. Snapshots of the projection of sites that have flipped at least once in the course of an annealing calculation at T = 0.8. For ease of visualization, we have only shown those spins that have flipped in three adjacent layers in a plane perpendicular to the surface. Each column shows a sequence of snapshots running in time from top to bottom separated by time intervals of 1.9 x $10^4$ (left column) and 1.1 x $10^5$ (right column). The response of two different films are shown (the 'mixed response' and 'front propagation' examples from Fig. 7). The left and right columns correspond to a films deposited at T = 0.5 and 0.3, respectively. Both films were deposited at the rate of $10^{-3}$.

Figure 12. The dependence of front velocities on the equilibrium bulk relaxation time for a range of annealing temperatures.

Figure 13. Plot of the spin, averaged over the whole film, evolving in time during annealing at T = 0.8. The crossover from the linear growth, diagnostic of front propagation, to stretched exponential relaxation, evidence of bulk relaxation, is evident as the fictive temperatures of the initial films increases.

**Appendix: The temperature dependence of the upper bound on the relaxation time for various surfaces of the 3spin facilitated model on the cubic lattice.**

The facilitated kinetic Ising model does allow us some flexibility in specifying the degree to which a free surface will enhance the kinetics in the adjacent liquid layer. Because of the underlying lattice used in the spin models, the orientation of the surface plane significantly changes the surface enhancement of kinetics. The 100, 110 and 111 planes of the cubic lattice correspond to surface planes in which sites have 1, 2 or 3 neighbours that are constrained to be all up, respectively. Depending on the choice of surface orientation, we can, therefore, produce free surfaces for the 3D – 3spin model which resemble a 2D – 2spin model (100), a set of uncoupled 1D – 1 spin models (110) or an array of sites without any kinetic constraint at all (111).

Before continuing we note that the discussion presented here concerning the kinetic variation between different free surfaces also pertains to different substrate bound surfaces. So, in order of an increasing constraint due to an all down substrate, we have the 100 surface being the least constrained, followed by the 110 and then the 111 glass-substrate interfaces.

As explained in the text, an upper bound on the relaxation time of the various free surfaces of the 3D-3spin facilitated kinetic Ising model is provided by spin models at reduced dimensions. In this section we collect expressions for the relaxation times associated with the various orientation of the free surface in the 3D model as well as obtaining an expression for the bulk relaxation time for the 3D-3spin model itself.

**A.1 The 111 Surface**

The 111 surface consists of isolated spins relaxing without constraint (i.e. via standard Glauber dynamics [38]) in the presence of an external field. As a consequence, the relaxation time $\tau_{0D}$ (where the '0D' notation is used for consistency with the other models treated here to reflect the fact that the uncoupled spins behave as independent points) is a constant, independent of temperature, reflecting the characteristic time scale of the elementary event, i.e.

$$\tau_{0D} = \text{constant} \tag{A1}$$

**A.2 The 110 Surface**

The 110 surface consists of a series of independent 1D-1spin facilitated models. Whitelam et al [39] have established that the relaxation time is proportional to $c^{-3}$. The final expression is

$$\tau_{1D} = \frac{\tau_{0D}}{\pi} c^{-3} \tag{A2}$$

where c, the equilibrium concentration of up spins, is provided by Eq. 1.

**A.3 The 100 Surface.**

The 2D-2spin facilitated kinetic Ising model provides the upper bound for the relaxation time for the 100 surface. The temperature dependence for the relaxation time $\tau_{2D}$ in this model has been determined previously [11]. A site was defined as 'active' if it was possible, by flipping up all of that spins flippable neighbours and then flipping up *their* flippable neighbours in turn and so on, one could eventually flip up all the spins in the system. The alternative is that the up flips come to an end due to the confinement of an enclosing 'wall' of down spins. The concentration of 'active' sites $Q_{2D}$ was found [11] to be given by

$$Q_{2D} = \left[6c^2(1-c)^2 + 4c^3(1-c) + c^4\right]\exp\left(\frac{(1-c)^8}{8\ln(1-c)}\right) \tag{A3}$$

It was then shown that the relaxation time was well described by a power law relation to $Q_{2D}$ of the form,

$$\tau_{2D} = \tau_{0D} 0.34 Q_{2D}^{-3.8}. \tag{A4}$$

**A.4 The 3D Bulk Relaxation Time**

Finally, we need an expression for the bulk relaxation time for the 3D-3spin model. Graham et al [23] have fitted the temperature dependence of this model to a Volger-Fulcher expression, $\tau \sim \exp(A/[T-0.68])$. The problem with such expression is that it cannot describe relaxation times at or below T=0.68 due to the divergence. This is a serious problem for the present study where we want to consider temperatures well below this value. Following a similar argument to that used to derive an expression for the concentration $Q_{2D}$ [11], we have derived the following expression for the concentration of active sites in the 3D-3spin model $Q_{3D}$,

$$Q_{3D} = q\exp\left[-6(1-c)^7\sqrt{\frac{-1}{24\ln(1-c)}}\left[1-\Phi(\sqrt{-6\ln(1-c)})\right]\right] \tag{A5}$$

where

$$q = 120c^3(1-c)^3 + 30c^4(1-c)^2 + 6c^5(1-c) + c^6 \tag{A6}$$

and $\Phi(x)$ is the error function,

$$\Phi(x) = \frac{2}{\sqrt{\pi}}\int_0^x dt e^{-t^2}$$

In Figure 1 we show the log-log plot of the relaxation time $\tau_{3D}$ plotted against $1/Q_{3D}$. We find the following power law relation,

$$\tau_{3D} = \tau_{0D} 31.77 Q_{3D}^{-3.95} \tag{A7}$$

The large exponent in Eq. A7 is very similar to that found in the 2D-2spin model [11].

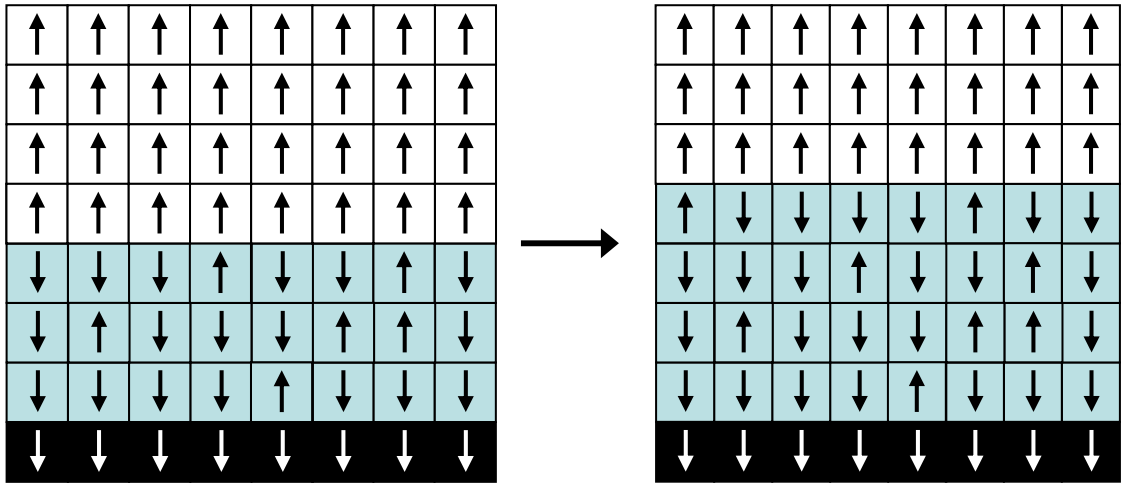

**Figure 1**.

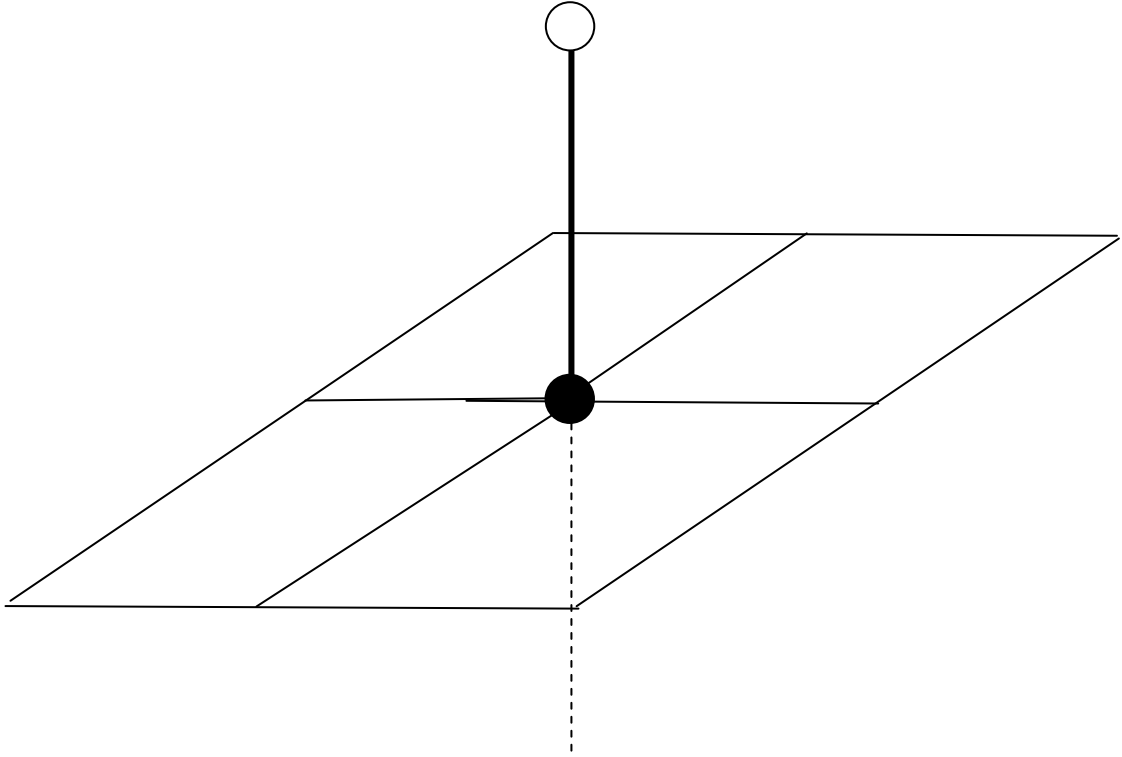

**Figure 2.**

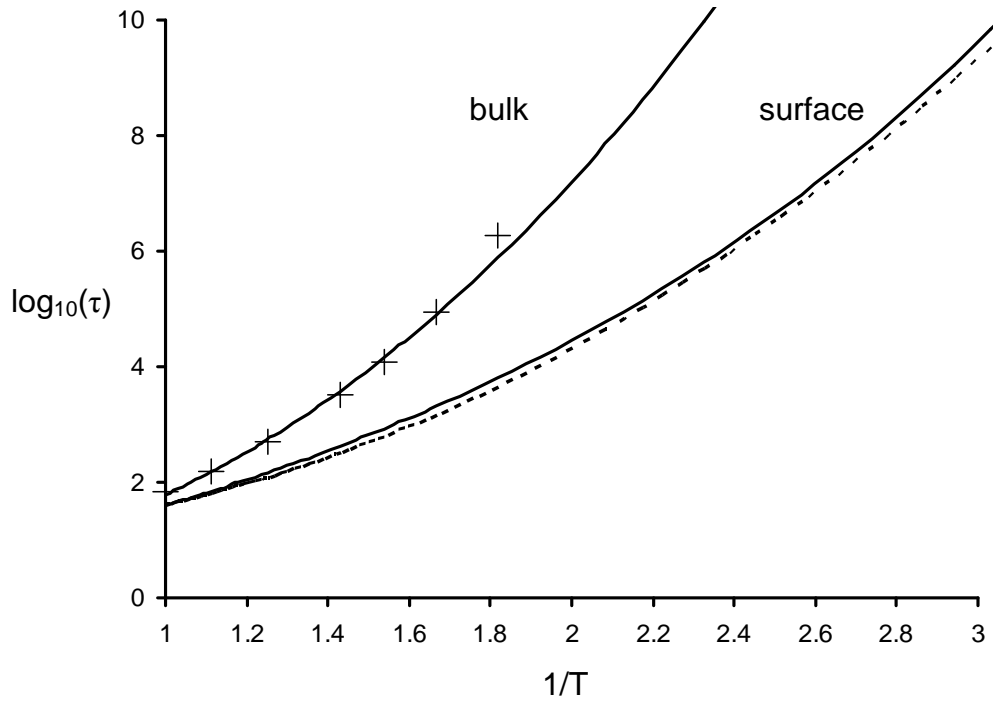

Figure 3

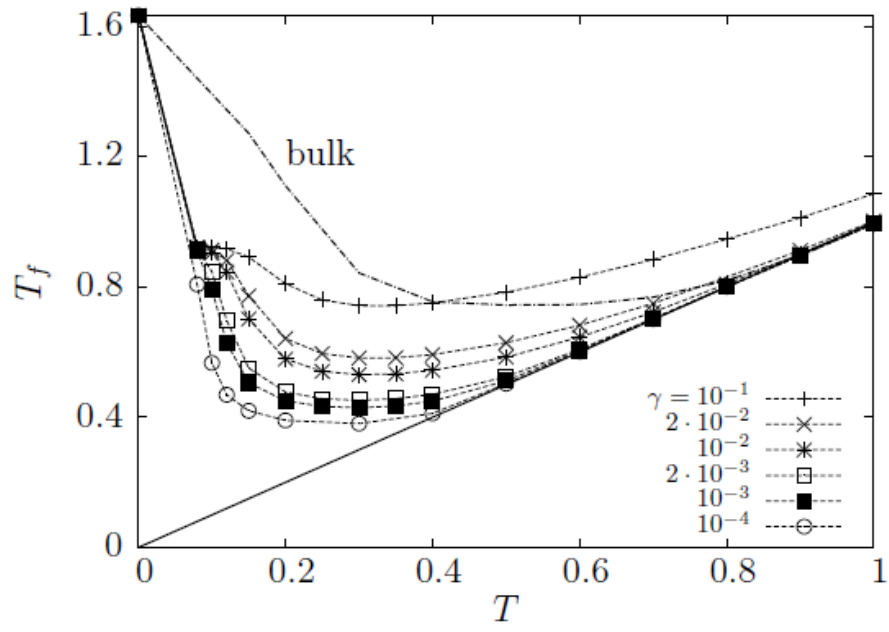

Figure 4

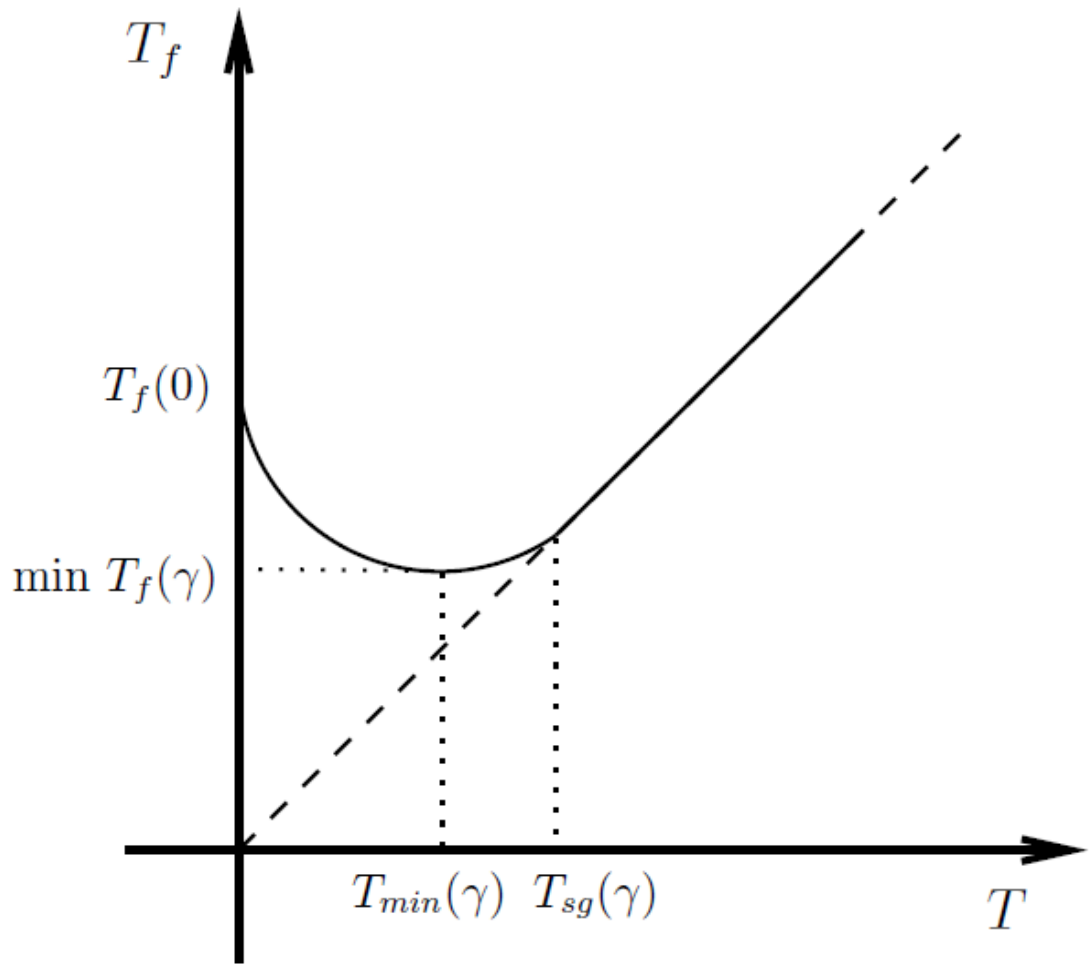

Figure 5

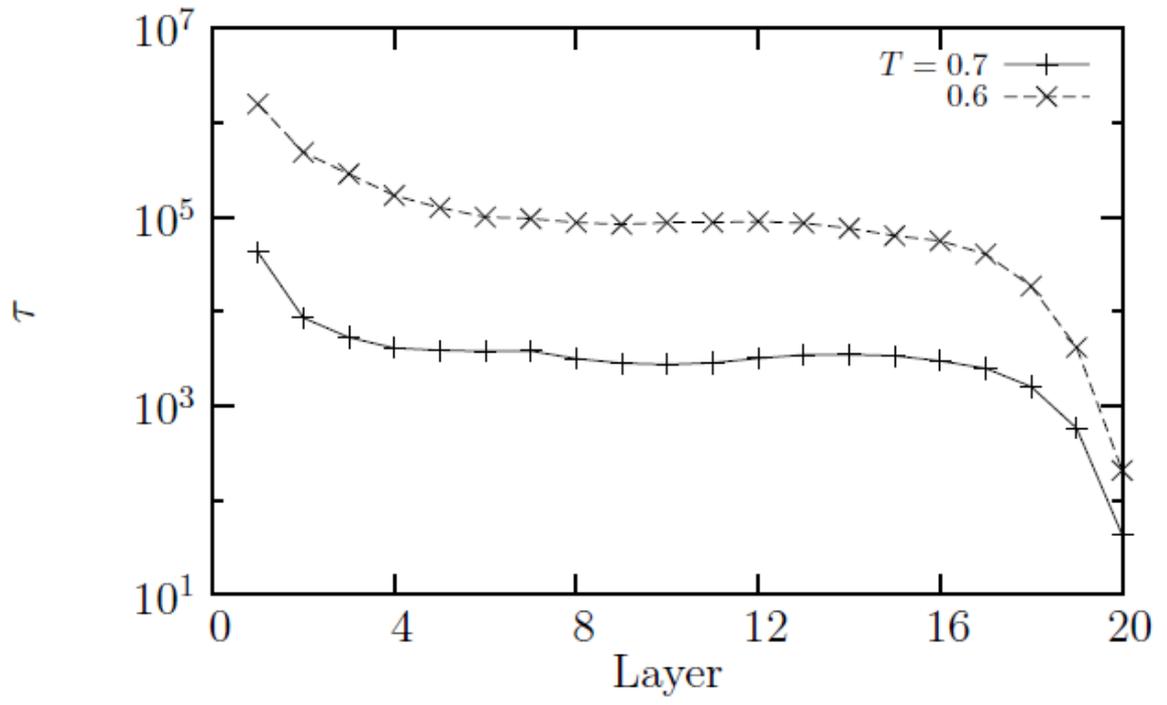

Figure 6

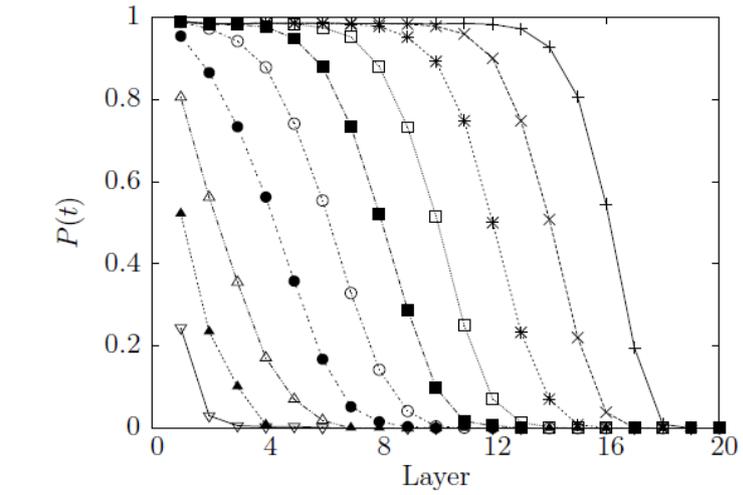

(a)

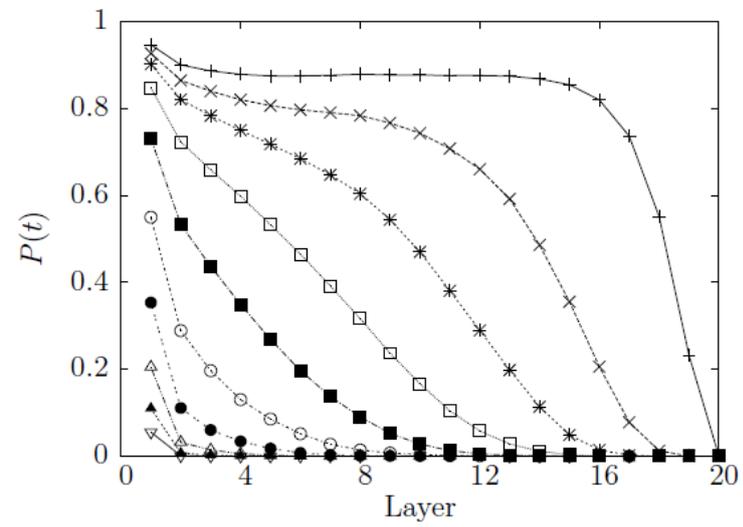

(b)

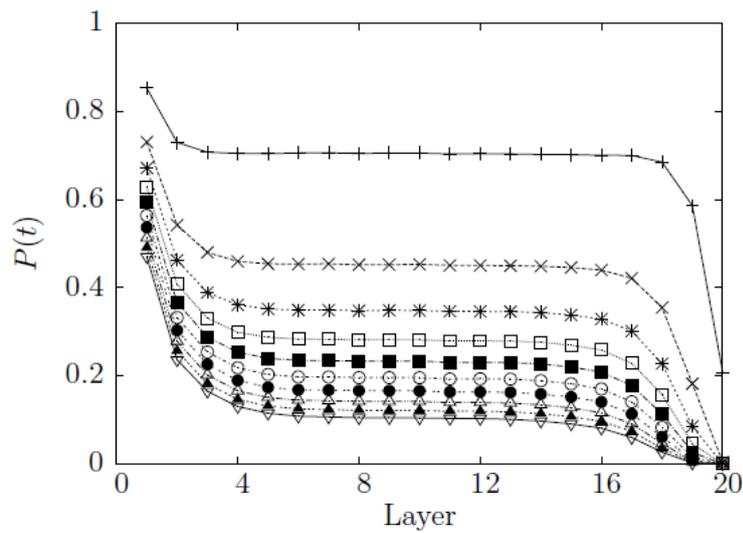

(c)

Figure 7

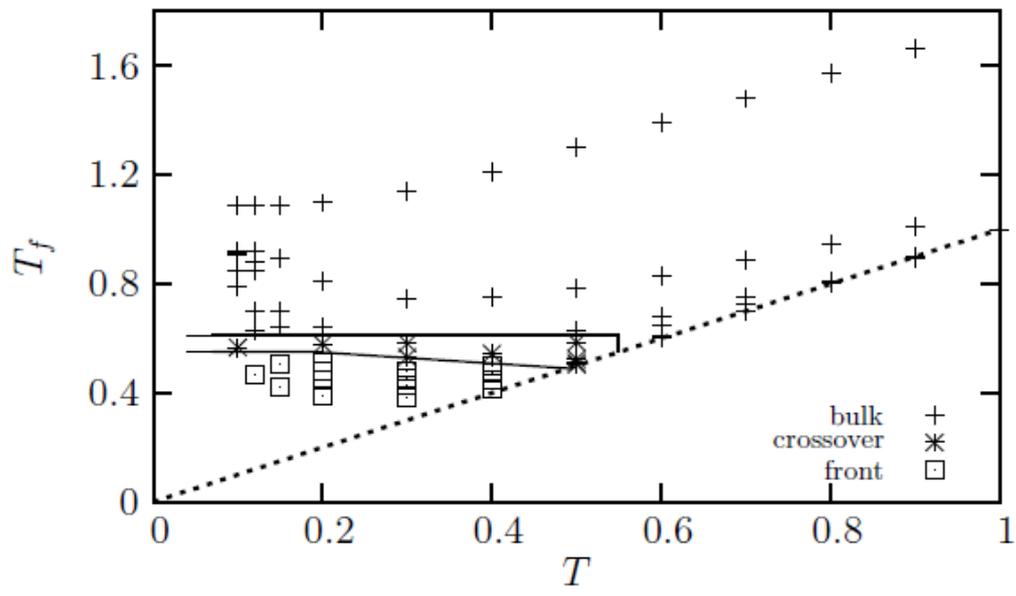

Figure 8

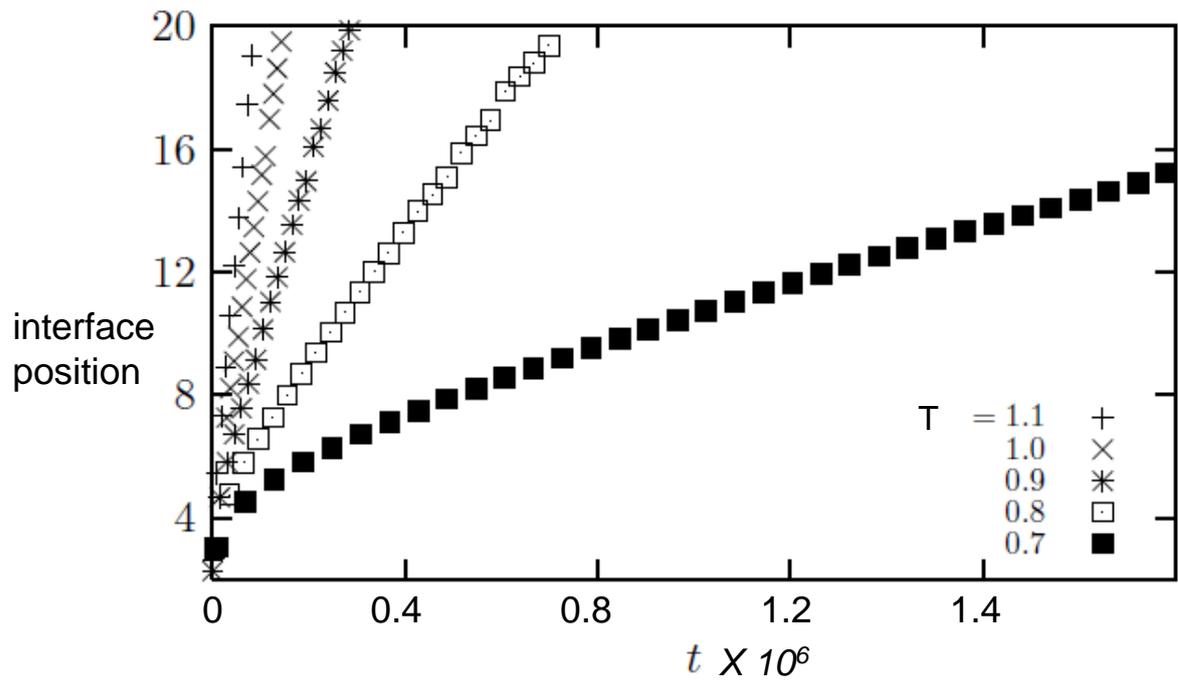

Figure 9

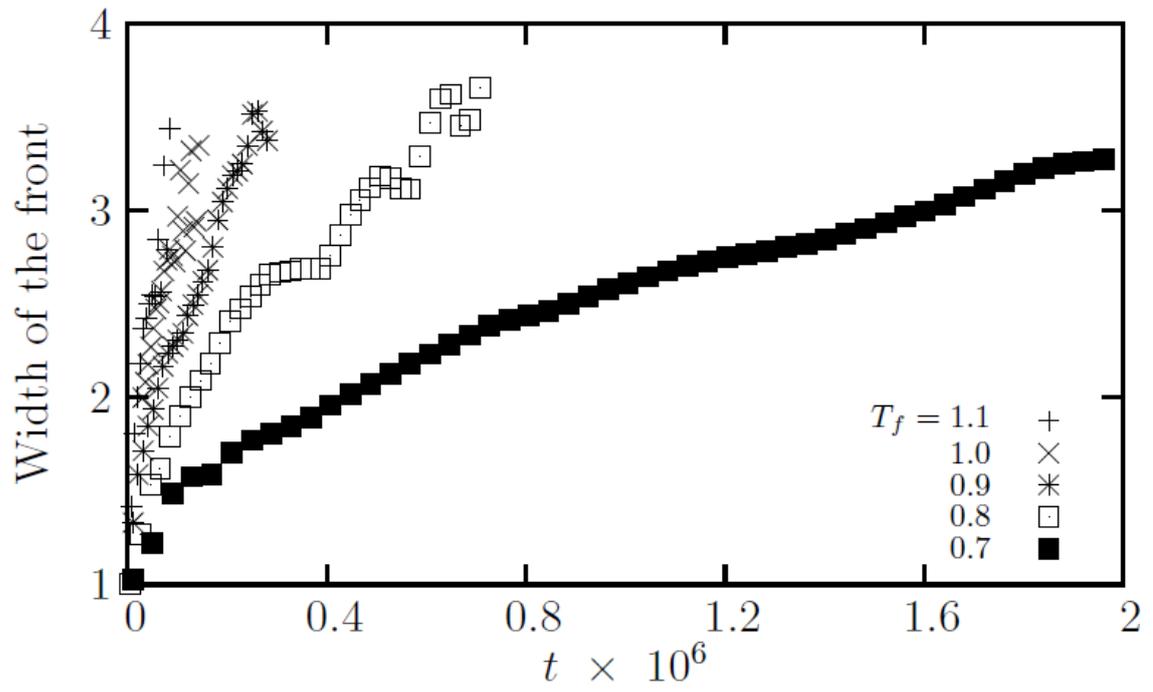

Figure 10

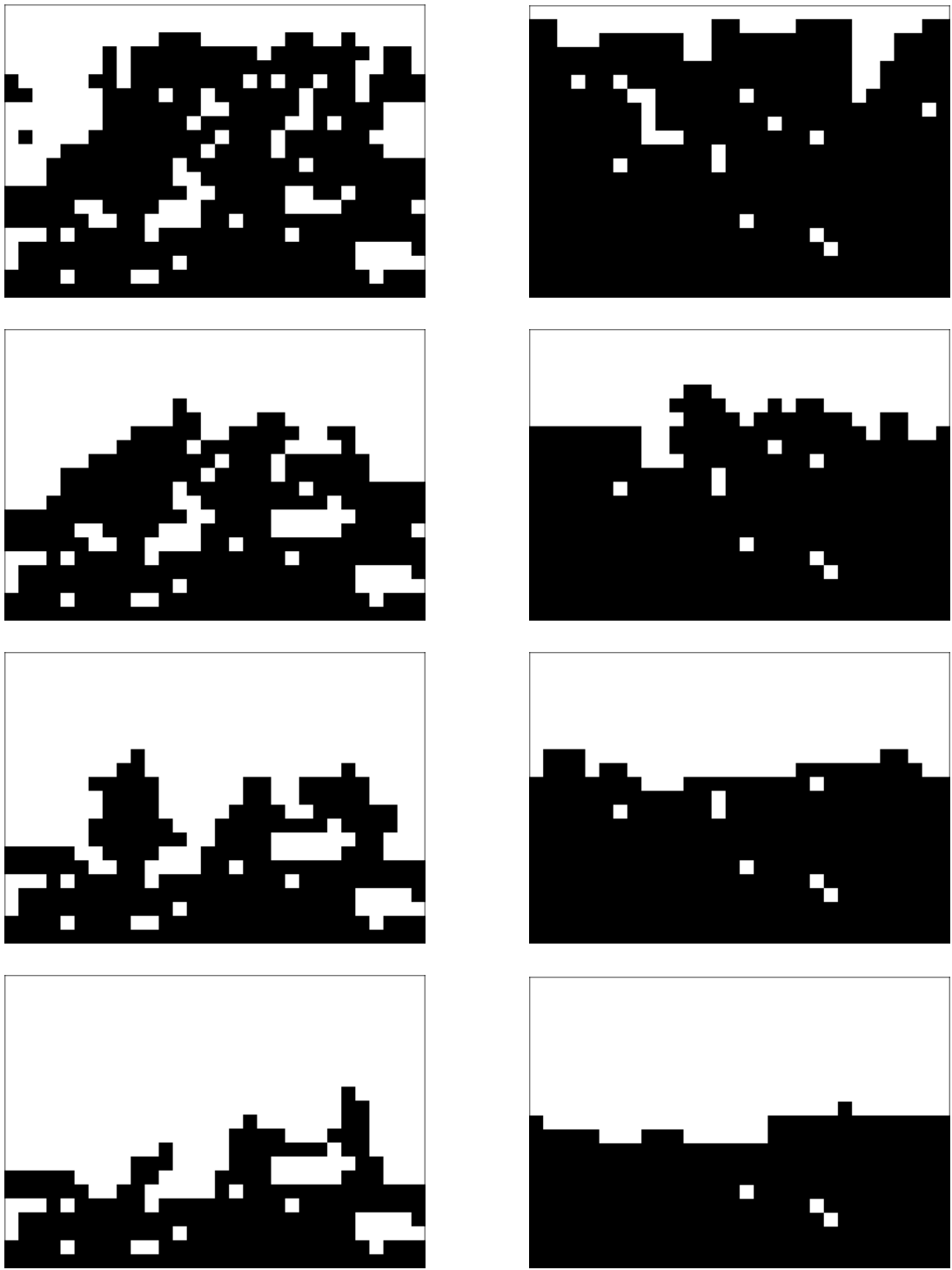

Figure 11

32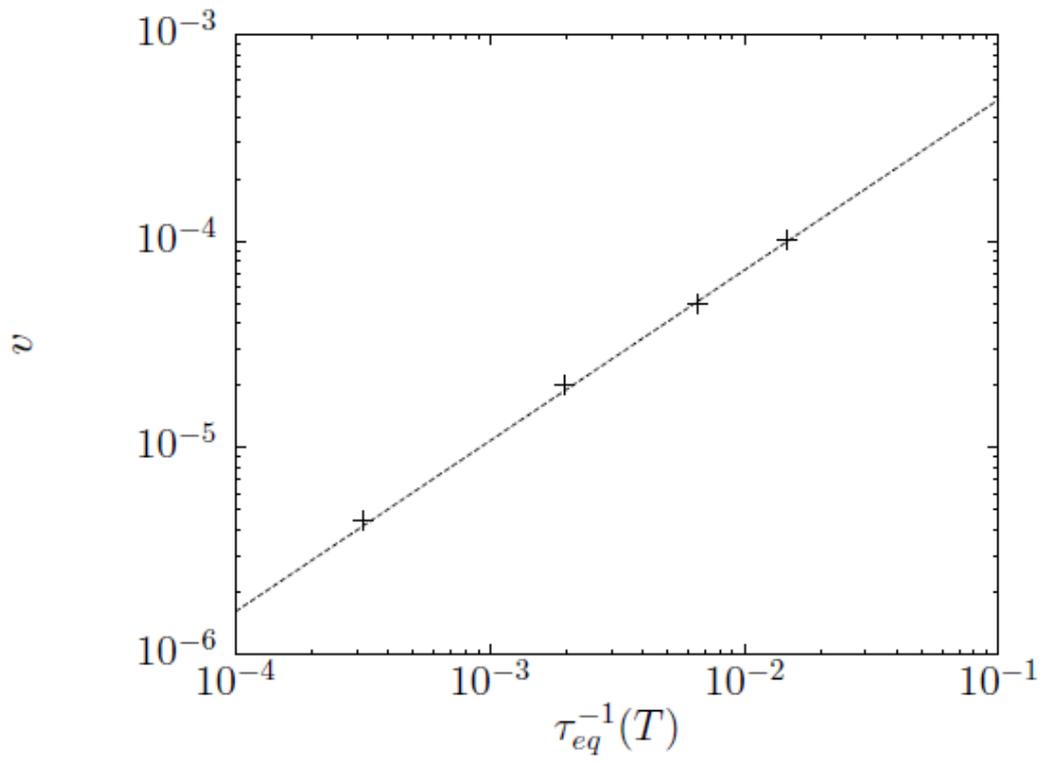

Figure 12

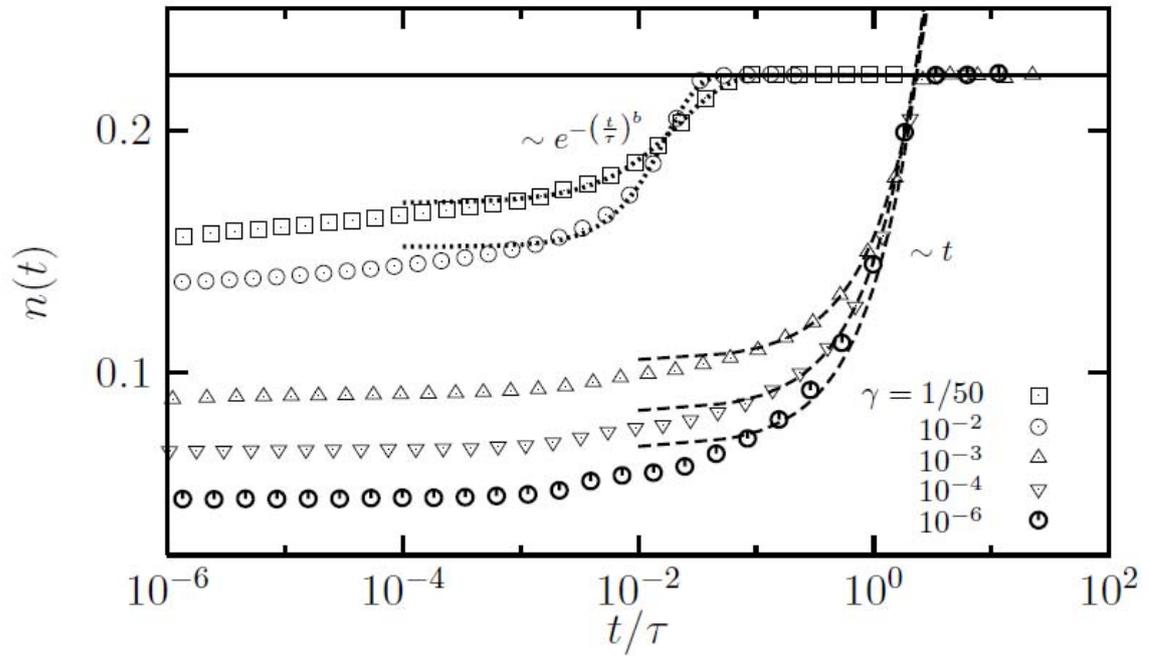

Figure 13